\documentclass[aip, cha, amsmath,amssymb,groupedaddress]{revtex4-1}
\usepackage{graphicx}
\usepackage{soul}
\usepackage{color}
\usepackage{amsmath}
\usepackage{pgf}
\usepackage[greek,english]{babel}

\begin{document}
\title{Flux bias-controlled chaos and extreme multistability in SQUID oscillators} 
\author{J. Hizanidis}
\author{N. Lazarides}
\author{G. P. Tsironis}
\affiliation{
 Department of Physics, University of Crete, P. O. Box 2208, 
 71003 Heraklion,Greece;  \\
National University of Science and Technology MISiS, Leninsky Prospect 4, 
Moscow, 119049, Russia
}
\date{\today}
\begin{abstract} 
The radio frequency (rf) Superconducting QUantum Interference Device (SQUID) is 
a highly nonlinear oscillator exhibiting rich dynamical behavior. It has been 
studied for many years and it has found numerous applications in magnetic field 
sensors, in biomagnetism, in non-destructive evaluation, and gradiometers, among
others. Despite its theoretical and practical importance, there is relatively 
very little work on its multistability, chaotic properties, and bifurcation 
structure. In the present work, the dynamical properties of the SQUID in the 
strongly nonlinear regime are demonstrated using a well-established model whose 
parameters lie in the experimentally accessible range of values.
When driven by a time-periodic (ac) flux 
either with or without a constant (dc) bias, the SQUID exhibits extreme 
multistability at frequencies around the (geometric) resonance. This effect is
manifested by a ``snake-like'' form of the resonance curve. In the presence of both ac and dc 
flux, multiple bifurcation sequences and secondary resonance branches appear at 
frequencies above and below the geometric resonance.
In the latter case, the SQUID exhibits
chaotic behavior in large regions of the parameter space; it is also found that 
the state of the SQUID can be switched from chaotic to periodic or vice versa 
by a slight variation of the dc flux.
\end{abstract}
\pacs{05.65.+b,05.45.Xt,78.67.Pt,89.75.-k,89.75.Kd}
\keywords{chaos, multistability, bifurcations}
\maketitle
{\bf In this work we study the dynamics of a nonlinear driven oscillator which 
serves as a model for a radio-frequency (rf) Superconducting QUantum 
Interference Device (SQUID), hereafter referred to as ``SQUID''. 
When arranged in periodic structures, SQUIDs form metamaterials with 
extraordinary electromagnetic properties and important quantum technology 
applications. Besides their appeal as superconducting devices, SQUID metamaterials
provide a unique testbed for exploring complex spatio-temporal dynamics. Their
properties are affected to a large extent by those of their constitutive elements.
Thus, an exploration of the dynamical properties of the single SQUID is
not only theoretically interesting but it may also help for the construction of 
SQUID-based devices and metamaterials with improved capabilities.
Here we revisit the single SQUID oscillator and undertake a systematic study of 
the flux bias effects on the system's dynamics. By using experimentally relevant 
parameters, we show that the single SQUID system can exhibit highly 
multistable responses to an externally applied driving force. This is reflected 
in the resonance curve which exhibits a ``snake-like'' structure around the 
geometric resonance frequency, while at lower frequencies a very delicate 
sequence of bifurcations leading to chaos is revealed, which we have quantified
using suitable measures. For the first time, we address in detail the role of an 
additional constant term in the external force driving the SQUID and show that 
it can induce new rich dynamics and complex bifurcation scenarios which are 
absent otherwise. The flux bias, which serves as yet another control parameter 
of the SQUID dynamics, has been largely ignored in prior works, and this is the 
focus of the interesting results presented in this manuscript.}

\section{Introduction}
\begin{figure}[h!]
 \includegraphics[width=.49\textwidth]{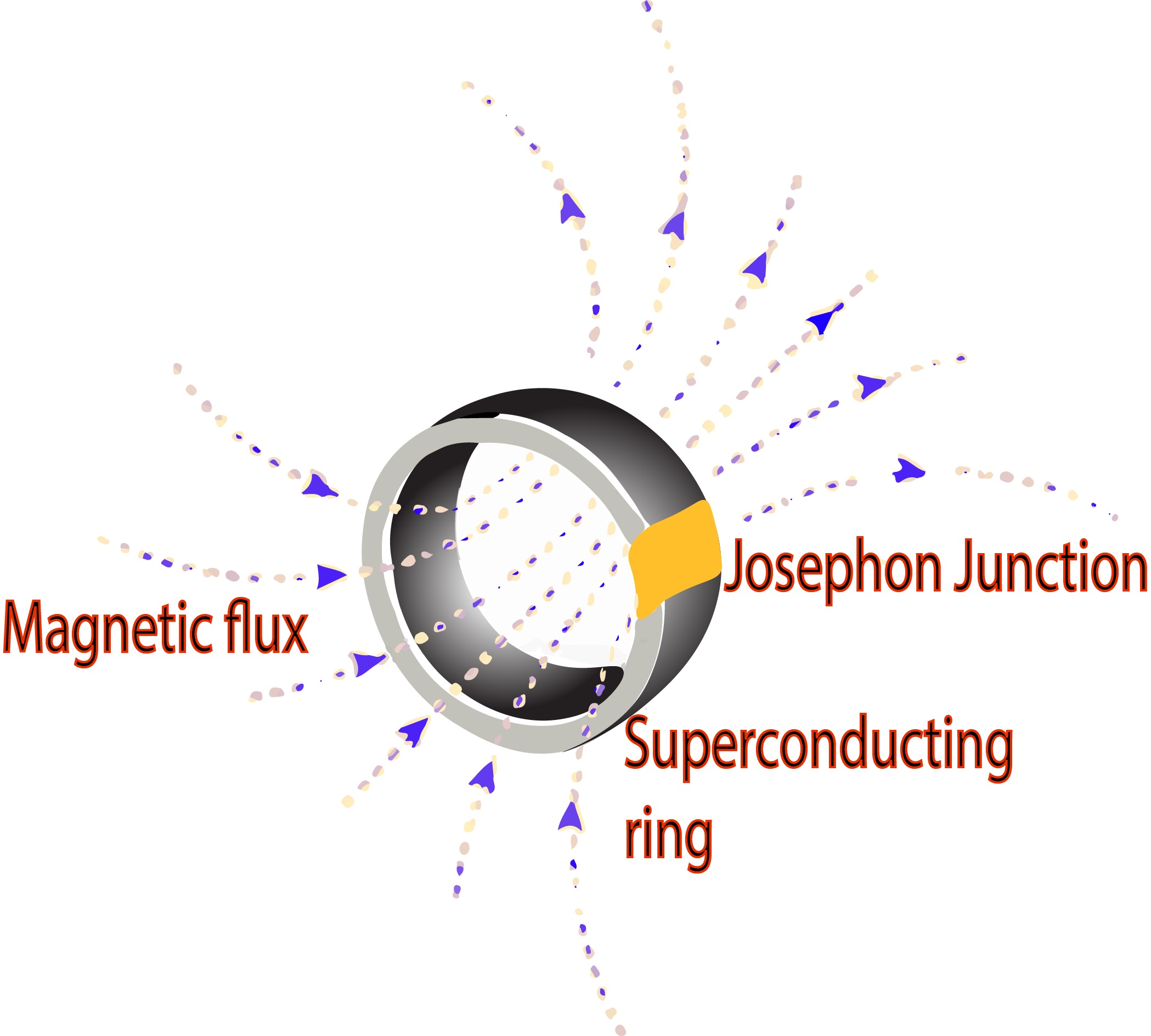}
 \includegraphics[width=.49\textwidth]{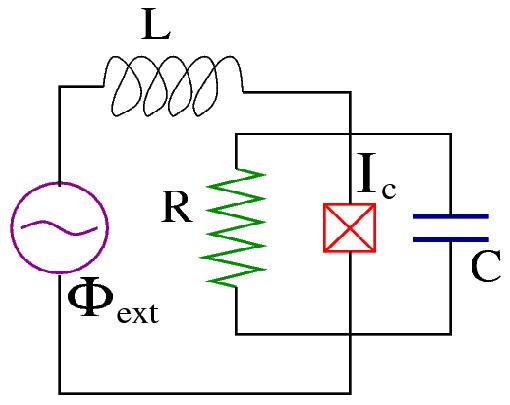}
 \caption{Schematic figure of a SQUID in a magnetic field (left) and its 
 equivalent electrical circuit (right).  
 \label{fig1}
}
\end{figure}

Superconducting metamaterials are artificially structured media of weakly 
coupled discrete elements that exhibit extraordinary properties 
\cite{Jung2014b,Ustinov2015,Lazarides2018}. Besides their main appeal as 
superconducting 
devices, these metamaterials provide a unique testbed for exploring complex 
spatio-temporal dynamics. The Superconducting Quantum Interference Device 
(SQUID) as a meta-atom (building block of a superconducting metamaterial) 
consists of a superconducting ring interrupted by a Josephson junction (JJ) 
as shown schematically in Fig.~\ref{fig1}. When driven by an alternating 
magnetic field, the induced supercurrents around the ring are determined 
through the celebrated Josephson relations~\cite{Likharev1986,Josephson1962}. 
Metamaterials comprising SQUIDs, i.~e., {\em SQUID metamaterials}, are very 
attractive for both theoretical and experimental research, holding a great 
promise for novel applications. A prominent feature of SQUID metamaterials 
is negative diamagnetic permeability that has been predicted both for the 
quantum \cite{Du2006} and the classical regime \cite{Lazarides2007}. Other 
properties of SQUID metamaterials include tunability of the resonance 
frequency with a flux bias \cite{Butz2013,Trepanier2013,Zhang2015}, dynamic 
multistability \cite{Jung2014a,Zhang2015}, and self-induced broadband 
transparency \cite{Zhang2015}. Recently, the collective behavior of coupled 
SQUID oscillators was also studied and counter-intuitive dynamic states
referred to as {\em chimera states} \cite{panaggio:2015,Kemeth2016} were 
found in a wide parameter regime \cite{LAZ15,HIZ16a,HIZ16b}.

Importantly, it has been shown that many of the properties of SQUID 
metamaterials emerge, to a large extent, from those of the individual SQUIDs, if 
all elements in the metamaterial are biased homogeneously and the coupling 
between them is weak \cite{Trepanier2013}. In particular, the key ingredient 
to the formation of the chimera states, i.~e., states of coexisting synchronized 
and unsynchronized clusters, is the multistability of the single SQUID. From the 
viewpoint of nonlinear dynamics, the SQUID is a strongly nonlinear oscillator 
with a resonant response to an externally applied alternating magnetic field; it 
exhibits rich dynamical behavior which we revisit in this article, revealing new 
interesting effects in a systematic way, which can be controlled by a flux bias 
term. 
 
The magnetic flux $\Phi$ threading the loop of the SQUID is given by:
\begin{eqnarray}
\label{eq1}
  \Phi =\Phi_{ext} +L\, I ,
\end{eqnarray}
where $\Phi_{ext}$ is the external flux applied to the SQUID, $L$ is the 
self-inductance of the SQUID ring, and
\begin{eqnarray}
\label{eq2}
  I =-C\frac{d^2\Phi}{dt^2} -\frac{1}{R} \frac{d\Phi}{dt} 
     -I_c\, \sin\left(2\pi\frac{\Phi}{\Phi_0}\right), 
\end{eqnarray}
is the current in the SQUID as provided by the resistively and capacitively 
shunted junction (RCSJ) model of the JJ \cite{Likharev1986}. In Eq. (\ref{eq2}),
within the RCSJ framework, $C$ is the capacitance of the JJ of the SQUID, $R$ is 
the resistance, $I_c$ is the critical current which characterizes the JJ, 
$\Phi_0$ is the flux quantum, and $t$ is the temporal variable. Combining Eqs. 
(\ref{eq1}) and (\ref{eq2}), gives the equation: 
\begin{eqnarray}
\label{eq2.2}
  C\frac{d^2\Phi}{dt^2} +\frac{1}{R} \frac{d\Phi}{dt} 
     +I_c\, \sin\left(2\pi\frac{\Phi}{\Phi_0}\right) +\frac{\Phi-\Phi_{ext}}{L}=0, 
\end{eqnarray}

which can be obtained from direct applications of Kirkhoff's laws to the 
equivalent electrical circuit for the SQUID (shown in Fig. \ref{fig1}).
The external flux takes the form: 
\begin{eqnarray}
\label{eq2.3}
  \Phi_{ext} =\Phi_{dc} +\Phi_{ac} \, \cos( \omega t ),
\end{eqnarray}
i.~e., it contains both constant (dc) flux bias $\Phi_{dc}$ and an alternating 
(ac) flux of amplitude $\Phi_{ac}$ and frequency $\omega$.
The normalized equation for the flux through the loop of the SQUID can
be obtained by combining Eqs. (\ref{eq2.2}) and (\ref{eq2.3}) and  
transforming the resulting equation using the relations:
\begin{eqnarray}
\label{eq2.4}
  \phi=\frac{\Phi}{\Phi_0}, ~~~\phi_{ac,dc}=\frac{\Phi_{ac,dc}}{\Phi_0},
  ~~~\tau=\frac{t}{\omega_{LC}^{-1}}, ~~~\Omega=\frac{\omega}{\omega_{LC}},
\end{eqnarray}
where $\omega_{LC} =1 / \sqrt{L C}$ is the inductive-capacitive SQUID frequency,
and the definitions
\begin{eqnarray}
\label{eq2.5}
   \beta=\frac{I_c L}{\Phi_0} =\frac{\beta_L}{2\pi}, \qquad
   \gamma=\frac{1}{R} \sqrt{ \frac{L}{C} }.
\end{eqnarray}
for the rescaled SQUID parameter and the loss coefficient, respectively.
Thus we get:
\begin{eqnarray}
\label{eq3}
   \ddot{\phi} +\gamma \dot{\phi} +\phi +\beta \sin\left( 2\pi \phi \right) =
\phi_{dc}+ \phi_{ac} \cos(\Omega \tau), 
\end{eqnarray}
which can be also written as 
\begin{eqnarray}
\label{eq3.2}
   \ddot{\phi} +\gamma \dot{\phi} =-\frac{d u_{SQ}}{d \phi},
\end{eqnarray}
where the normalized SQUID potential reads:
\begin{eqnarray}
\label{eq3.3}
   u_{SQ} =-\phi_{ext} (\tau) \phi 
    +\frac{1}{2} \left( \phi^2 -\frac{\beta}{\pi} \cos(2\pi \phi) \right),
\end{eqnarray}
and the normalized external flux is given by:
\begin{eqnarray}
\label{eq3.4}
   \phi_{ext} (\tau) =\phi_{dc} +\phi_{ac} \cos( \Omega \tau ).
\end{eqnarray}

\begin{figure}[h!]
 \includegraphics[width=\textwidth]{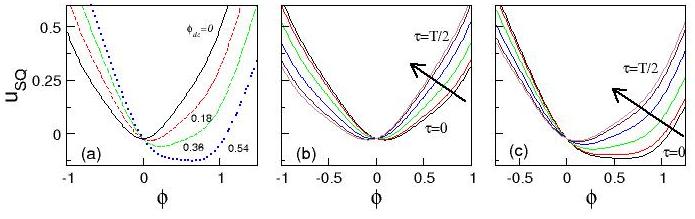}
 \caption{
 The SQUID potential $u_{SQ}$ from Eq. (\ref{eq3.3}) for $\beta=0.1369$ 
 ($\beta_L \simeq 0.86$).
 (a) $u_{SQ} (\phi)$ for $\phi_{ac}=0$ and $\phi_{dc} =0$ (black-solid), $0.18$ (red-dashed),
$0.36$ (green-long-dashed), $0.54$ (blue-dotted).
 (b) Temporal evolution of $u_{SQ} (\phi)$ during half of the driving period 
 $T=2\pi/\Omega$, for $\phi_{ac} =0.16$, $\beta=0.1369$ 
 ($\beta_L \simeq 0.86$), $\Omega=0.345$, and $\phi_{dc} =0$. 
 (c) Similar to (b) for $\phi_{dc} =0.36$. The arrows point the direction of the
 time increasing from $\tau=0$ to $\tau=T/2$.	
 \label{fig2}
}
\end{figure}
The SQUID potential $u_{SQ}$, given by Eq. (\ref{eq3.3}), becomes time-dependent 
for $\phi_{ac} \neq 0$. For $\phi_{ac}=0$, although the potential is constant in time, its
shape changes with $\phi_{dc}$; more specifically, while it is symmetric for 
$\phi_{dc}=0$, it becomes more and more asymmetric with increasing $\phi_{dc}$,
as can be observed in Fig. \ref{fig2}(a). For $\phi_{ac} \neq 0$, several 
snapshots of the time-dependent $u_{SQ}$ are shown in Figs. \ref{fig2}(b) and 
(c), for $\phi_{dc}=0$ and $\phi_{dc}=0.36$, respectively, during the first half
of the driving period $T=2\pi/\Omega$.

The SQUID model has been studied in the past in the hysteretic regime 
($\beta_L>1$), where low dimensional chaos was reported for varying ac flux 
\cite{FES83,RIT83,RIT84,SOE85}. The effect of noise has also been  
studied in the SQUID system, in particular with respect to stochastic resonance~\cite{GAM98}.
Moreover, in Refs. \cite{SCH88,BRU88,JAC89} the 
Melnikov method was applied to the SQUID in the case of small ac field and the 
threshold conditions for the onset of homoclinic behavior leading to chaos were 
found. In all of these works, the dc component of the external magnetic flux was 
set to zero. The role of $\phi_{dc}$ on the SQUID dynamics has been investigated 
in \cite{BUL86,BUL91} but only through an analytical approximation of the SQUID 
equation. The effect of a dc flux bias on the 
dynamics of a hysteretic SQUID has been also discussed recently in Ref. 
\cite{CAP12}. In particular, it was demonstrated theoretically that the state of
the system can be shifted from one fixed point to another via the dc flux.   
In the present work, we revisit the dynamics of the full model for a 
SQUID oscillator in the non-hysteretic regime ($\beta_L <1$) and reveal the 
complex behavior induced by all parameters of the flux bias, with emphasis on 
the effect of the dc term. 
Our work contributes significantly to recent experimental findings, 
where a number of dynamic properties of single, non-hysteretic SQUIDs were demonstrated,
such as multistability, switching \cite{Jung2014a},
and broad-band tunability of the resonance 
frequency by magnetic fields and temperature
\cite{Butz2013,Trepanier2013,Zhang2015}.

\begin{figure}[h!]
\includegraphics[width=0.9\textwidth]{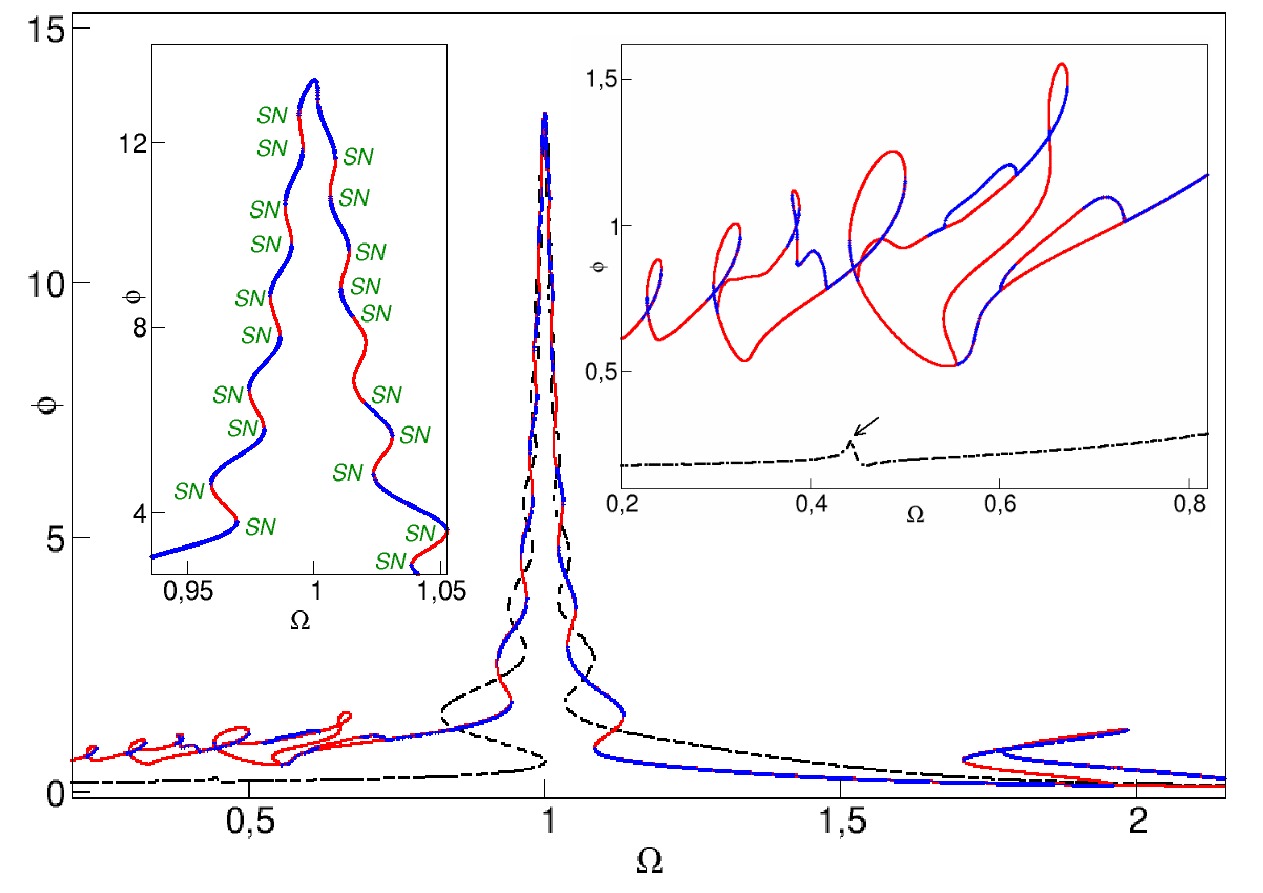}
\caption{
 The ``snake-like'' resonance curve of the SQUID for $\beta=0.1369$ 
 ($\beta_L = 0.86$), $\gamma=0.024$, external ac flux amplitude $\phi_{ac}=0.16$ 
 and dc flux $\phi_{dc}=0.36$.
 Blue and red lines correspond to branches of stable and unstable periodic 
 solutions, respectively. The black dashed curve corresponds to $\phi_{dc}=0$. 
 Insets: Enlargement around the maximum multistability frequency (left) and 
 around lower frequencies (right).
 The symbols ``SN'' in the left inset denote saddle-node bifurcations of limit 
 cycles.
\label{fig3}
}
\end{figure}
The design parameters of a SQUID are its self-inductance $L$, the capacitance $C$
of the JJ, the critical current $I_c$ of the JJ, and the subgap resistance 
$R$. Typical values of these parameters are $L=120 ~pH$, $C=1.1 ~pF$, 
$I_c =2.35 ~\mu A$, and $R=500 ~\Omega$ \cite{Trepanier2013,Zhang2015}. These 
parameters provide the values of the dimensionless coefficients 
$\beta \simeq 0.1369$ ($\beta_L \simeq 0,86$) and $\gamma \simeq 0.024$ which 
appear in the normalized Eq. (\ref{eq3}) for the flux $\phi =\Phi / \Phi_0$ 
through the loop of the SQUID. They also provide $f_{LC} =\omega_{LC} /(2\pi) \simeq 13.9 ~GHz$ 
($\Omega \simeq 1$) and $f_{SQ} =\omega_{SQ} /(2\pi) \simeq 18.9 ~GHz$ 
($\Omega =\Omega_{SQ} \simeq 1.364$) for the geometric and the linear 
resonance frequency of the SQUID, respectively, which are also typical in 
single-SQUID experiments \cite{Butz2013,Trepanier2013,Zhang2015}. The values of 
the externally controlled parameters $\phi_{dc}$, $\phi_{ac}$, and $\Omega$ used 
here, are within the range of the experimentally accessible values, i.~e., 
$\phi_{dc}$ in the interval $[-1, 2]$ \cite{Trepanier2013}, $\phi_{ac}$ in the 
interval $[0.001, 0.18]$ \cite{Zhang2015}, and $\Omega$ in the interval 
$\frac{2\pi}{\omega_{LC}} [10, 22.5] ~GHz$ \cite{Trepanier2013}.

\section{Snake-like resonance curve and subresonances}
As mentioned previously, in the present work we focus on the non-hysteretic regime 
of our system. In the context of SQUID dynamics, the terms hysteretic and non-hysteretic refer to
its static properties. In practice, the potential $u_{SQ}$ of a non-hysteretic
SQUID has only one minimum (as shown in Fig. \ref{fig2}), while the 
corresponding potential $u_{SQ}$ of a hysteretic SQUID has more than one, which 
gives rise to multistability. In the case of a single-well $u_{SQ}$ (which is the case here), the multistability
is not related to the minima of the potential, but instead it is a purely dynamical effect. In the 
linear regime, i.~e., for $|\phi_{ext} (\tau)| \ll 1$, the flux amplitude-frequency (resonance) 
curve of the SQUID resembles that of a harmonic oscillator 
with eigenfrequency $\Omega_{SQ}$. However, with increasing $\phi_{ac}$ and/or 
$\phi_{dc}$, nonlinear effects become more and more significant. The resonance
curve bends more and more; in the strongly nonlinear regime, it acquires a 
``snake-like'' form with several stable and unstable branches.   
By setting $\gamma=0$ and $\phi_{ext}=0$ into Eq. (\ref{eq3}), linearizing,
and using $\beta \sin( 2\pi \phi) \simeq \beta_L \phi$, the linear resonance 
frequency of the SQUID can be obtained as $\Omega_{SQ} =\sqrt{1 +\beta_L}$.
Note that the {\em geometric} resonance frequency is $\Omega =1$ 
(or $\omega =\omega_{LC}$ in natural units), which is always lower than 
$\Omega_{SQ}$. The resonance frequency of the SQUID can be tuned by varying
$\phi_{ac}$ and/or $\phi_{dc}$; thus that frequnecy can be shifted from 
$\Omega \simeq \Omega_{SQ}$ in the linear regime, to $\Omega \simeq 1$ in the 
strongly nonlinear regime. It is the latter regime that is investigated here.
When the SQUID is driven by an ac flux, the flux through the loop of the SQUID
oscillates with a particular amplitude; its frequency or equivalently its period 
of oscillation is that of the driving flux (although oscillations with periods
several times that of the driving flux or even chaotic oscillations can be also
observed in the strongly nonlinear regime, as we shall see below).

In this Section, we present in detail the delicate structure of the SQUID 
resonance curve in a wide range of values for the driving frequency $\Omega$. 
Figure \ref{fig3} shows the amplitude of the flux oscillations calculated from 
Eq.~(\ref{eq3}) over the frequency of the ac flux field $\Omega$ for finite 
ac and dc flux. This curve has been obtained using a very powerful software tool
that executes a root-finding algorithm for continuation of steady state solutions and 
bifurcation problems~\cite{ENG02}. This tool also allows us to determine the stability of 
the system's periodic solutions through the calculation of the corresponding Floquet multipliers.
In relation to stability, in Fig.~\ref{fig3}, the blue and red branches mark the stable and unstable 
periodic (period-1) solutions, respectively. This ``snake-like'' form of the 
resonance curve has also been observed in the Duffing equation~\cite{HIG90,KOV08}.
Notably, a snaking resonance curve for a nonlinear superconducting quantum
oscillator which is very similar to the one shown in Fig.~\ref{fig3} has
been reported in figure 3 of Ref. \cite{HRI11}. The curve depicts the 
number of photons absorbed for the occurrence of {\em quantum phase slips} 
(which are the cause of nonlinearity) as a function of the driving frequency.
It could be argued that this curve is the dual quantum analogue of 
the snaking resonance curves presented here.
In our model the ``snake-like'' resonance curve was first reported in the context of chimera states 
\cite{HIZ16b}, where its crucial role for the emergence of these patterns was 
discussed. In Ref.~\cite{HIZ16b} the dc flux $\phi_{dc}$ was set to zero, 
which in Fig.~\ref{fig3} corresponds to the black dashed line. Both curves 
exhibit a winding behavior around the geometric resonance frequency 
$\omega_{LC}$, where multiple saddle-node bifurcations of limit cycles take 
place when stable and unstable branches merge (marked by ``SN'' in the left inset).

\begin{figure}
\includegraphics[width=0.9\textwidth]{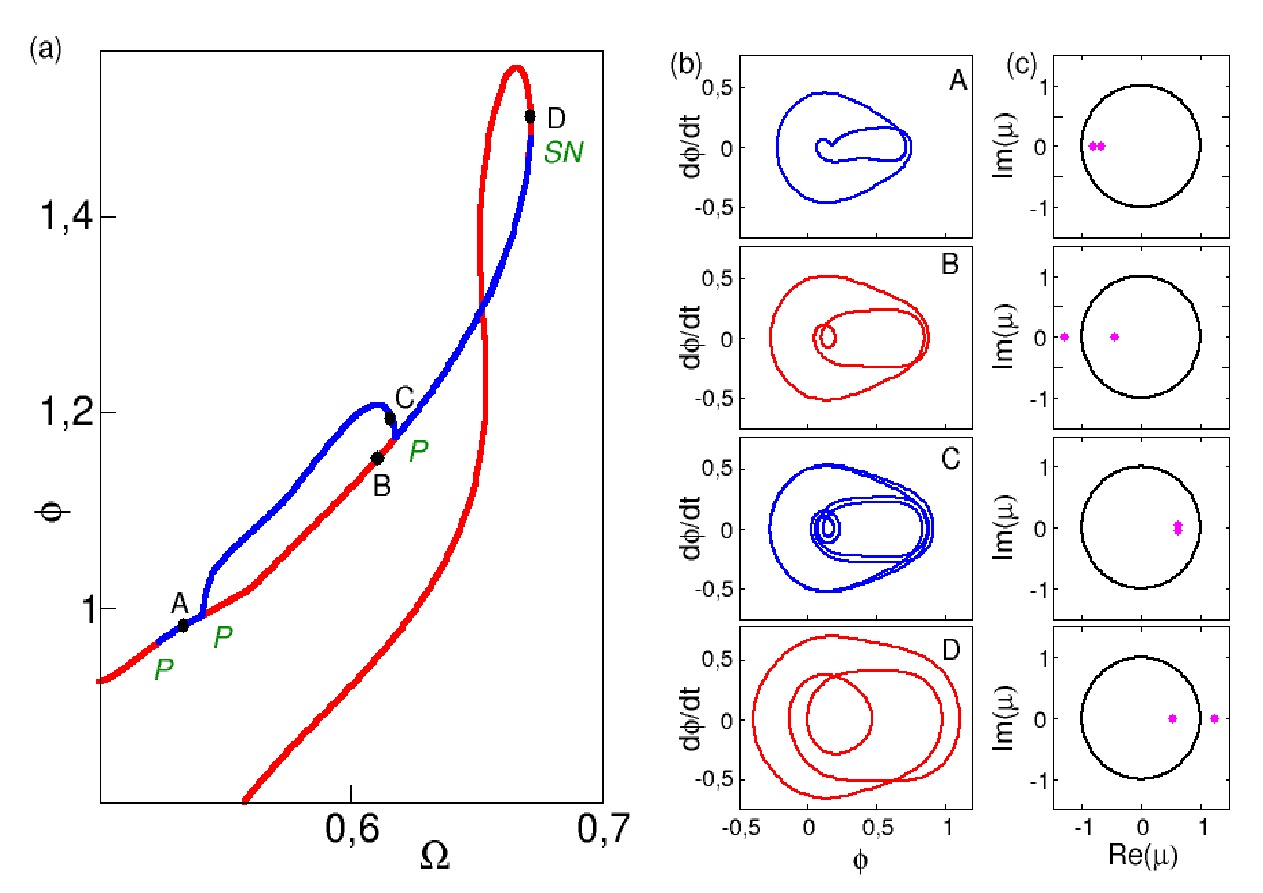}
\caption{
(a) Blowup of the resonance curve of Fig.~\ref{fig3}. Blue and red lines 
    correspond to branches of stable and unstable periodic solutions. ``SN'' and 
    ``P'' denote saddle-node and period doubling bifurcations, respectively.
(b) Phase portraits of the periodic orbits of points $A-D$ on the resonance 
    curve, and 
(c) corresponding leading Floquet multipliers in the complex unit circle.  
    Other parameters as in Fig.~\ref{fig3}.
\label{fig4}
}
\end{figure}
However, the inclusion of a dc component in the external magnetic flux creates 
some new effects at lower and secondarily at higher frequencies. As shown by the 
right inset, at lower frequencies, $\phi_{dc}$ induces new stable and unstable 
periodic solutions which create a complex structure of branches which are absent 
for zero dc flux. We can distinguish multiple secondary maxima one of which 
appears around $\Omega=0.45$ like in the zero dc flux curve. This value is about 
one third of the SQUID linear resonance frequency.

The other subresonances are shifted to lower values compared to other fractions 
of $\Omega_{SQ}$ and this is due to the nonlinearity which is more prominent for 
finite $\phi_{dc}$. This is related to the fact that the SQUID potential becomes more and more
asymmetric as $\phi_{dc}$ increases (see Fig.~\ref{fig2}) enhancing, thus, the nonlinear effects 
in our system. The bifurcation structure in the low frequency regime is very 
complex and one example is shown in the enlargement in Fig.~\ref{fig4}(a). 
We can see that in this small interval of $\Omega$, four bifurcations take place: 
One saddle-node (``SN'') and three period-doubling bifurcations denoted by the 
letter ``P''. In Fig.~\ref{fig4}(b) we plot the phase portraits of the periodic 
solutions corresponding to points $A-D$ of Fig.~\ref{fig4}(a). The orbit 
corresponding to $A$ is a stable period-2 solution which becomes unstable at the 
second period-doubling bifurcation. This is evident by the corresponding Floquet 
multipliers (Fig.~\ref{fig4}(c)) whose real part cross the complex unit circle 
through $-1$. Simultaneously, a new stable period-4 orbit is created belonging 
to the stable branch which emerges at the ``P'' bifurcation point after point $A$. 
Its phase portrait for $\Omega=0.6167$ corresponds to point $C$ and is shown 
in Fig.~\ref{fig4}(b) along with its Floquet multipliers in Fig.~\ref{fig4}(c) 
which, as expected, lie in the complex unit circle. As $\Omega$ increases 
further, the real part of the Floquet multipliers exit the complex unit circle 
through $+1$ and a saddle-node bifurcation (``SN'') takes place, giving birth to 
an unstable orbit (point $D$).

\section{Bifurcation diagrams and attractors}
\begin{figure*}[!t]
  \includegraphics[angle=0,width=0.3\textwidth]{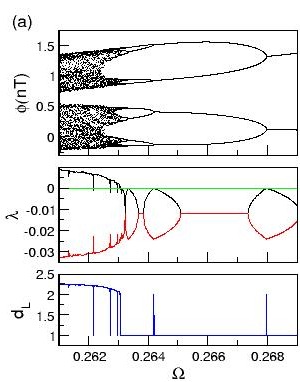} 
  \hspace{3mm}   
  \includegraphics[angle=0,width=0.3\textwidth]{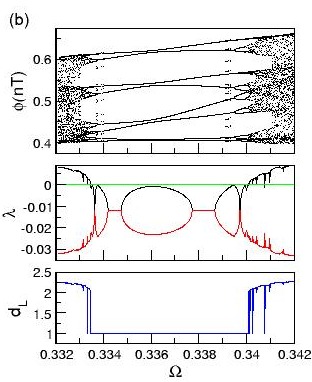} 
  \includegraphics[angle=0,width=0.3\textwidth]{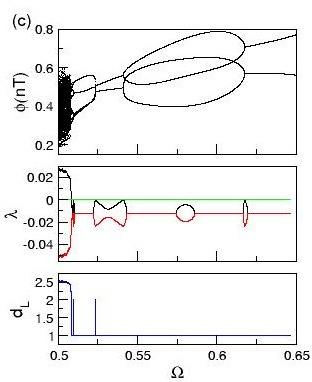} 
\caption{
 Dynamical behavior of a SQUID subjected to an external flux having both ac 
 and dc components (Eq. (\ref{eq3.4})) as a function of the driving frequency 
 $\Omega$ for three different intervals.
 For all three subfigures:
 Upper panel: bifurcation diagram. 
 Middle panel: all three Lyapunov exponents. 
 Lower panel: the corresponding fractal (Kaplan-Yorke or Lyapunov) dimension $d_L$.
 Other parameters as in Fig.~\ref{fig3}.
\label{fig5}
}
\end{figure*}

The sequence of bifurcations described above, takes place on the respective 
branches of periodic solutions. These coexist with multiple other branches 
which can be found by continuation from different initial conditions. Moreover, 
the period-doubling bifurcations discussed in Fig.~\ref{fig4} may lead the 
system to chaotic dynamics. Therefore, the full bifurcation diagram with 
$\Omega$ as the control parameter is much more complex, and parts of it can be 
seen in the upper panels of Fig.~\ref{fig5}: In Fig. \ref{fig5}(a) we observe 
a typical period-doubling bifurcation cascade which leads to chaos at 
$\Omega \simeq 0.263$. In Fig. \ref{fig5}(b) we observe a period-doubling 
bifurcation cascade of a period-8 solution which is followed by a reverse such 
cascade. In the upper panel of Fig.~\ref{fig5}(c), besides the period-doubling 
cascade at low frequencies $\Omega$, we observe the formation of a stable 
period-4 ``bubble''~\cite{BIE84}. The latter is created (destroyed) through a period-doubling (reverse 
period-doubling) bifurcation at $\Omega \simeq 0.54$ ($\Omega \simeq 0.62$). 
The appearance of bubbles here is a manifestation 
of ``antimonotonicity'', i.~e., the concurrent creation and destruction of periodic 
orbits, which has been observed in several physical systems such as driven 
nonlinear $RLC$ circuits \cite{NEW96}.
The above described bifurcation diagrams have been
produced via direct numerical integration of our model equation (Eq.~\ref{eq3}) 
and continuation in $\Omega$, for different initial condition realizations.

Next, we proceed to the quantification of the 
chaotic orbits through suitable measures. There are several ways to define the 
fractal dimension of a chaotic attractor, which is a measure of its geometric 
scaling properties or its ``complexity'' and it has been considered its most basic 
property. These methods fall into two categories, those derived from the 
topology, and those derived from the dynamics. Perhaps the most common of the 
former metrics is the correlation dimension~\cite{GRA83} and the most common of 
the latter type is the Lyapunov dimension $d_L$, proposed by Kaplan and Yorke
\cite{KAP79}. According to the definition of Kaplan and Yorke,
\begin{equation}
\label{01}
  d_L = k +\frac{1}{|\lambda_{k+1}|} \sum_{i=1}^k \lambda_i,
\end{equation}
where the Lyapunov exponents $\lambda_i$ are calculated from the system's equations
of motion using the method described 
in~\cite{WOL85} and $k$ is defined by the condition that
\begin{equation}
\label{02}
  \sum_{i=1}^k \lambda_i \geq 0 ~~~{\rm and} ~~~ \sum_{i=1}^{k+1} \lambda_i < 0.
\end{equation}
For example, the Lyapunov dimension of the chaotic attractor for $\Omega=0.5$
shown in the upper-right panel of Fig. \ref{fig5}(a) is $d_L \simeq 2.53$. 
In Figs. \ref{fig5}(a)-(c), we see a clear one-to-one correspondence between the
bifurcation plots and the curves of the Lyapunov exponents. That is, chaotic
dynamics emerges when the largest Lyapunov exponent becomes positive, while the
bifurcation points correspond to its zeroing. In the lower panel of each 
subfigure, the calculated fractal (Lyapunov or Kaplan-Yorke) dimension of the 
phase space attractors of the SQUID are also shown. Note that the Lyapunov 
dimension $d_L$ of chaotic attractors is always between $2$ and $3$ as it should 
be, i.~e., the chaotic phase space attractor is topologically complicated more 
than a limit cycle and less than a three-dimensional object. The geometric 
complexity of phase space attractors can be compared through their Lyapunov 
dimension.

\begin{figure*}
\includegraphics[width=1\textwidth]{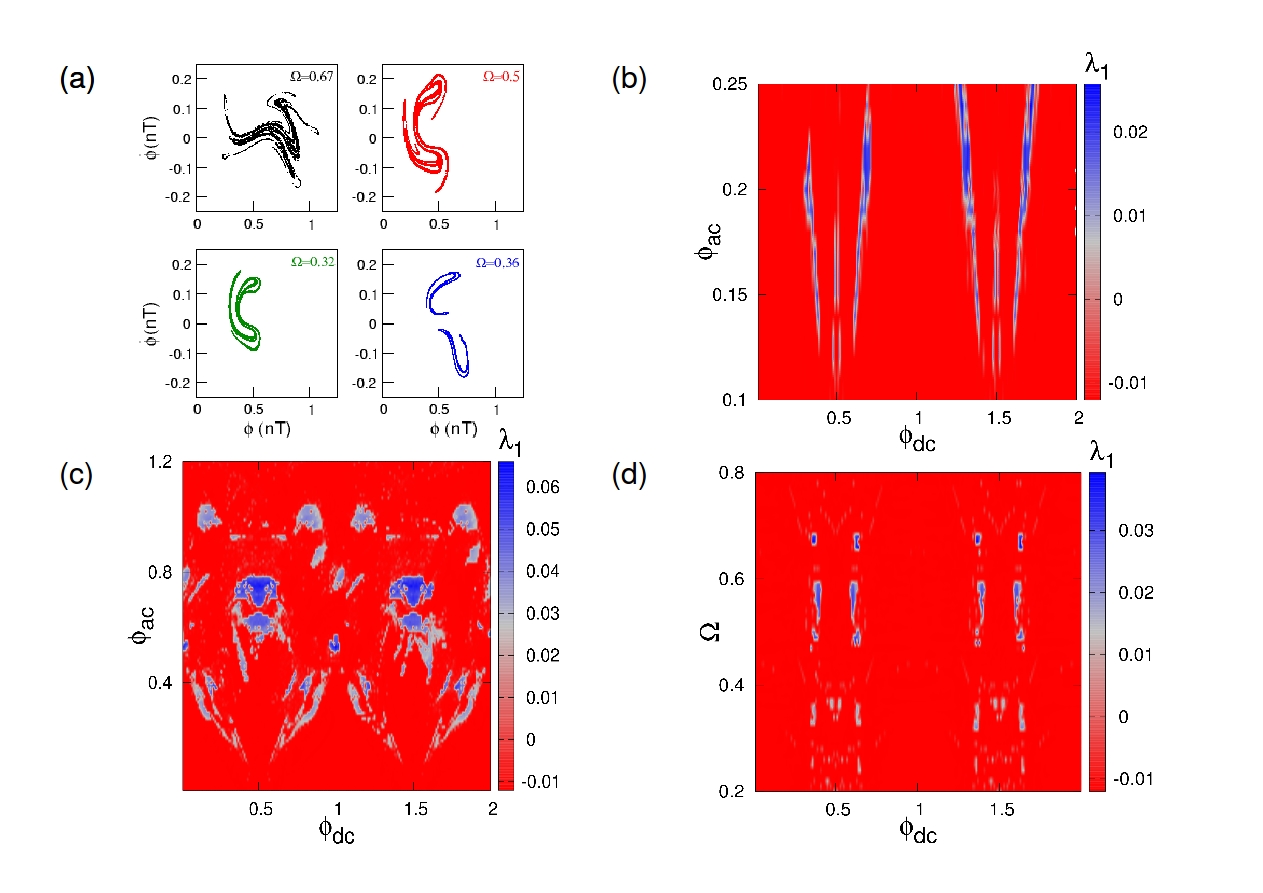}
\caption{
(a) Stroboscopic maps of chaotic attractors on the $\phi - \dot{\phi}$ phase 
    space for four different frequencies $\Omega$ indicated in each subfigure, 
    for $\phi_{ac} =0.16$, and $\phi_{dc} =0.36$. 
(b) The maximum Lyapunov exponent $\lambda_1$ mapped onto the 
    $(\phi_{dc},\phi_{ac})$ plane for $\Omega=0.345$, and 
(c) $\Omega=0.51$. 
(d) The maximum Lyapunov exponent $\lambda_1$ mapped onto the $(\phi_{dc},\Omega)$ 
    plane for $\phi_{ac}=0.16$.
Other parameters as in Fig.~\ref{fig3}.
\label{fig6}}
\end{figure*}

In order to observe the complexity of these phase-space attractors, we have also 
obtained the stroboscopic maps (as the complete phase space attractors are not 
very easy to visualize). In Fig. \ref{fig6}(a), four such attractors are shown 
for four different values of the driving frequency $\Omega$ (indicated on each 
subfigure).

\begin{figure}
\includegraphics[width=0.9\textwidth]{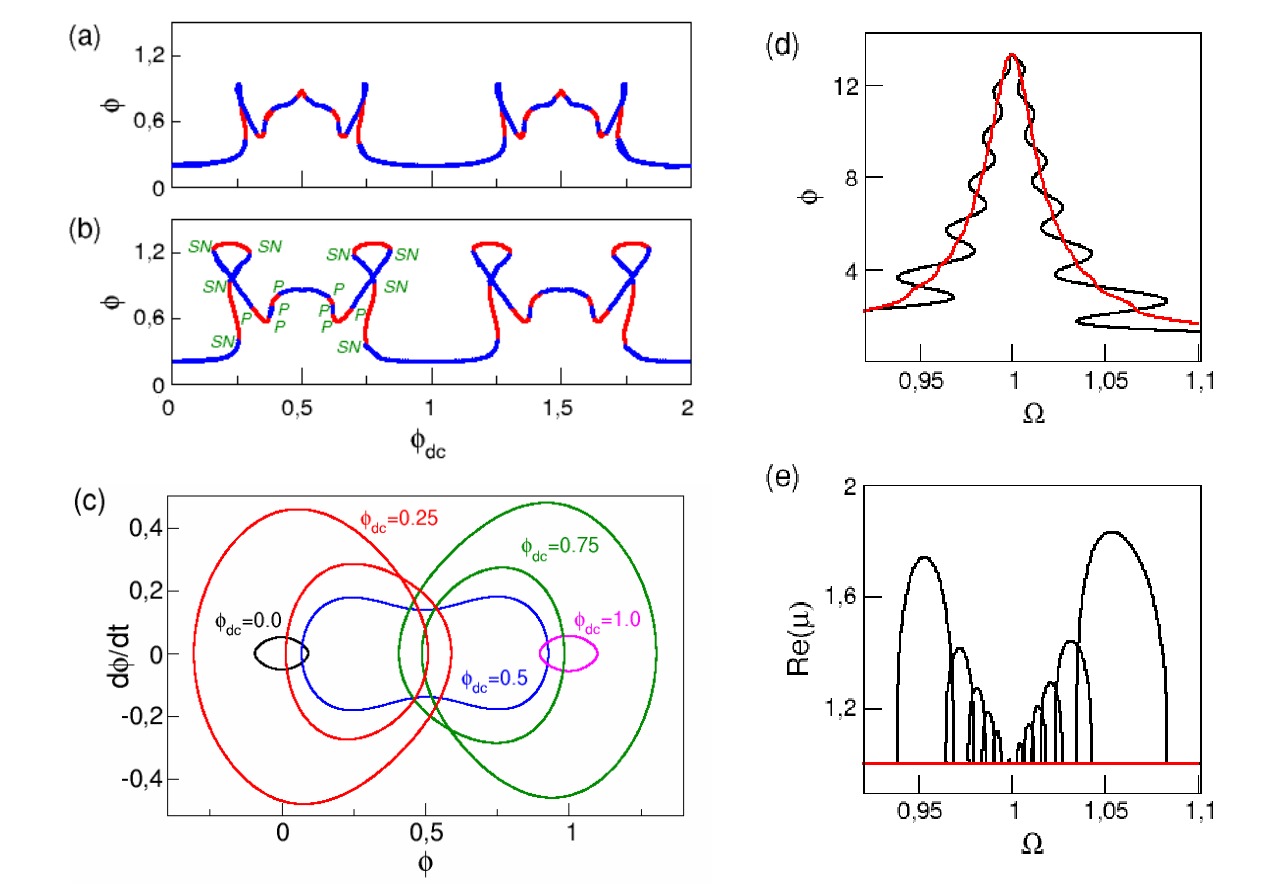}
\caption{
 Amplitude of the magnetic flux $\phi$ in dependence of the dc flux for two 
 different values of the external frequency 
(a) $\Omega=0.345$ and (b) $\Omega=0.51$.
 Blue and red lines correspond to branches of stable and unstable periodic 
 solutions, while ``SN'' and ``P'' in (b) denote saddle-node and period doubling 
 bifurcations, respectively.
(c) Phase portraits of the magnetic flux for five different dc fluxes and 
 $\Omega=0.51$.
(d) Resonance curve for $\phi_{dc}=0.10$ (black) and $\phi_{dc}=0.25$ (red) and 
(e) real part of the corresponding leading Floquet multiplier.
Other parameters as in Fig.~\ref{fig3}.
\label{fig7}}
\end{figure}

Chaotic behavior arises by varying all three parameters of the external magnetic 
flux. This is depicted in Figs.~\ref{fig6}(b)-(d), where the largest Lyapunov 
exponent $\lambda_1$ is calculated in the relevant parameter spaces. In general, 
chaos is observed for moderate to higher values of $\phi_{ac}$ 
(Figs.~\ref{fig6}(b), (c)) and for $\Omega$ values lower than $0.7$. Note that 
the chaotic regimes show approximately periodic structure in dependence of the 
dc flux $\phi_{dc}$. In order to explore this further, we take a cross section 
of Figs.~\ref{fig6}(b) and (c) at $\phi_{ac}=0.16$ and follow a single branch of 
periodic solutions up to $\phi_{dc}=1.0$. These are depicted in the left panels 
of Fig.~\ref{fig7}(a) and (b) where again, blue and red denote stable and 
unstable periodic solutions and in Fig.~\ref{fig7}(b) we have also noted the 
bifurcations that occur along the branch: There is a cascade of eight period 
doubling bifurcations preceded and followed by a sequence of four saddle-node 
bifurcations, which is repeated periodically for $\phi_{dc}>1$. The periodicity 
with the dc flux was discussed in reference \cite{Trepanier2013}, where an 
analytical expression of the solution of the linearized version of Eq.~\ref{eq3} 
was found as a function of $\phi_{dc}$. Note that there is also a symmetry in 
the bifurcation points, around $\phi_{dc}=0.5$, which is also reflected in 
Fig.~\ref{fig7}(c) where the phase portraits for various $\phi_{dc}$ values 
are shown for $\Omega=0.51$: The orbit for $\phi_{dc}=0.5$ lies in the center 
and around it we have two symmetrical period-1 orbits of low amplitude for 
$\phi_{dc}=0.0, 1.0$ and two symmetrical period-2 orbits of higher amplitude for 
$\phi_{dc}=0.25, 0.75$.

The external flux bias $\phi_{dc}$ certainly affects the ``snake-like'' resonance 
curve of the SQUID as well. First of all, the frequency $\Omega$ at which the first 
saddle-node bifurcation emerges fluctuates periodically with $\phi_{dc}$ (not shown here).
Moreover, around the resonance, the saddle-node bifurcations change from
sub- to supercritical. Most importantly, for sufficiently 
high values of $\phi_{ac}>0.12$, there are particular values of $\phi_{dc}$ at 
which those saddle-node bifurcations disappear! This is evident in 
Fig.~\ref{fig7}(d), in which the resonance curve is shown for two values of 
$\phi_{dc}$. We can see that for $\phi_{dc}=0.25$ the saddle-node bifurcations 
vanish and the real part of the corresponding Floquet multipliers never crosses 
the value $+1$ (Fig.~\ref{fig7}(e)). The same holds periodically for $\phi_{dc} =0.75, 1.25, \dots$ (not 
shown here). Note that for relatively low values of $\phi_{ac}$ the saddle-node
bifurcations do not disappear for any value of $\phi_{dc}$. 

\begin{figure}
\includegraphics[width=0.9\textwidth]{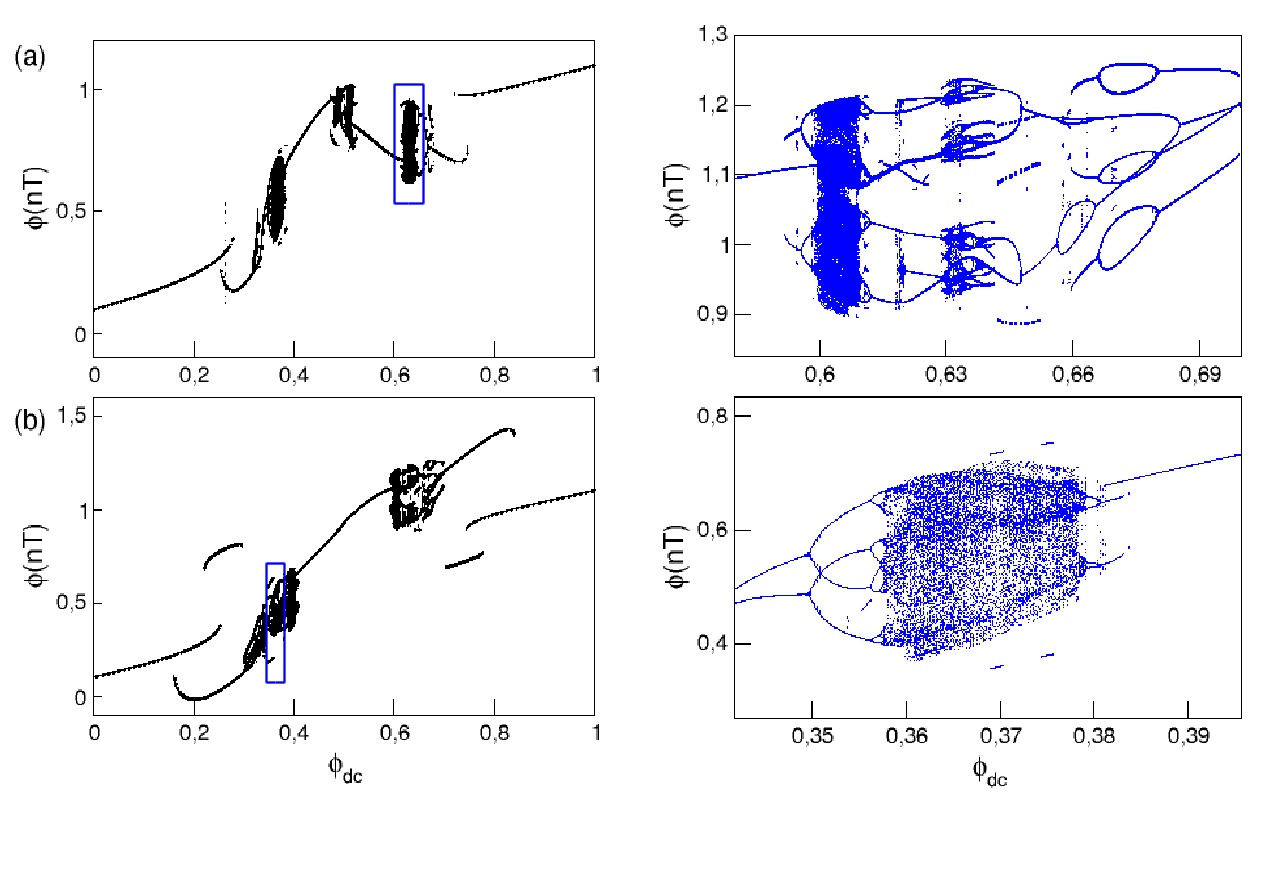}
\vspace{-1cm} \\ 
\caption{ 
 Amplitude of the magnetic flux recorded at integer multiples of the external 
 period (left panel) for (a) $\Omega=0.345$ and (b) $\Omega=0.51$ and 
 corresponding blowups (right panel).
 Other parameters as in Fig.~\ref{fig3}.
\label{fig8}}
\end{figure}

In addition to the effects described above, $\phi_{dc}$ can create multiple 
periodic solution branches which coexist and may lead to chaos through complex 
bifurcation scenarios. This is depicted in Fig.~\ref{fig8}, where the bifurcation 
diagrams for two different values of $\Omega$ are plotted. In the enlargements 
of the right panel of (a), we can see two period-doubling cascades coexisting 
and leading to a common chaotic regime which ends in eight branches of periodic 
solutions. Another interesting feature is the formation of four period-2 
``bubbles'' for higher values of $\phi_{dc}$. In the enlargement of (b) we see a 
similar period doubling cascade as in (a), but here the two branches lie within 
one-another. 

\section{Conclusions}
The dynamic properties of an rf SQUID, which is a highly nonlinear oscillator,
were explored numerically using the well-established model Eq. (\ref{eq3}) 
governing the temporal evolution of the flux $\phi (\tau)$ through the loop of 
the SQUID. The coefficients entering the model equation have been obtained from
relevant experimental parameters, and thus our results can be in principle
experimentally testable. The SQUID was subjected to the sum of a constant and a
sinusoidal flux. It is demonstrated that a non-zero flux bias crucially affects 
the dynamics of the SQUID in a wide range of driving flux amplitudes and 
frequencies.

At frequencies around the geometrical resonance $\omega_{LC}$, in particular, 
the flux bias $\phi_{dc}$ changes the snaking resonance curve by varying the 
location of the saddle-node 
bifurcations while it transforms them from subcritical to supercritical or vice 
versa. At frequencies higher than $\omega_{LC}$, it strongly 
enhances secondary resonances, which may also exhibit multistability similar to
that of the primary resonance curve. At frequencies lower than $\omega_{LC}$,
wide-band chaotic behavior has been observed for reasonable and experimentally 
accessible values of the flux bias $\phi_{dc}$ and the driving amplitude 
$\phi_{ac}$. A wealth of nonlinear dynamics effects such as period-doubling and 
reverse period-doubling, multi-periodic solutions, saddle-node bifurcations, 
bubbling and multistability, has been observed in this region. It should be noted that 
no period-doubling bifurcation cascades and 
subsequent transitions to chaos have been observed for $\phi_{dc}=0$.
This is probably due to the symmetry of time-independent part of the 
SQUID potential in the absence of a constant flux bias, which renders 
the SQUID equation (Eq.~\ref{eq3}) symmetric according to the considerations in 
Ref. \cite{SWI84}. That symmetry suppresses period-doubling 
bifurcations in a large class of systems, including the sinusoidally
driven damped pendulum \cite{SWI84}. The bifurcation diagrams with varying $\phi_{dc}$ at 
frequencies lower than $\omega_{LC}$ clearly reveal multistability, where 
periodic/multi-periodic and chaotic solutions may coexist. This allows for 
switching the SQUID dynamics from periodic to multi-periodic to chaotic with 
a slight variation of $\phi_{dc}$.

Extensive calculations of the maximal Lyapunov exponent in two external 
parameter spaces indicates that wide-band chaotic behavior appears at 
frequencies lower than $\omega_{LC}$ and relatively high $\phi_{dc}$ and 
$\phi_{ac}$. However, some regions of $\phi_{dc}$ and $\phi_{ac}$ are 
still experimentally accessible. Note that for relatively high $\phi_{ac}$,
the resonance curve may loose its ``snake-like'' form for a certain value of 
$\phi_{dc}=0.25$.  The vanishing of the resonance curve ``snake-like'' form   
is repeated periodically in the dc fluc and marks the
switching from chaotic to regular behavior. This study can prove
very useful for the deeper understanding of
the collective dynamics of coupled SQUIDs which form metametarials that find
important technological applications. Moreover, it may initiate further 
experimental work on the dynamics of single SQUIDs to confirm the predictions
above.

\subsection*{Acknowledgement}
This work was supported by 
the Ministry of Education and Science of the Russian Federation in the framework 
of the Increase Competitiveness Program of NUST ``MISiS'' (Grant No. К3-2017-057).
J.~H. is thankful to the General Secretariat for Research and Technology (GSRT)
for the financial support through the Postdoctoral Researchers Projects Fund. 



\end{document}